\documentclass{elsart}
\usepackage[dvips]{graphics}

\begin{document}

\begin{frontmatter}

\title{ACTIVE GALACTIC NUCLEI AS HIGH ENERGY ENGINES}
\vskip 20pt
\author{\bf by G. Henri$^{1,2}$, G. Pelletier$^{1,2}$, P.O. Petrucci$^1$,
N. Renaud$^1$ }
\address{1\ Laboratoire d'Astrophysique de l'Observatoire de Grenoble \\
2\ Institut Universitaire de France}
\maketitle
\vskip 12pt
\begin{abstract}
Active Galactic Nuclei are considered as possible sites of cosmic ray acceleration
and some of them have been observed as high energy gamma ray emitters
(Blazars). There naturally comes an appealing idea that the
acceleration of the highest energy cosmic rays in the AGNs has a
signature in the form of gamma ray emission and high energy neutrino
emission through the collisions of very high energy protons with soft
photons. Moreover it is often said that electrons cannot reach enough
energy through Fermi acceleration to account for the highest energy photons
observed with ground Cerenkov telescopes. In this paper, we discussed
these points and show that the fast variability of the flares recently
observed rules out the assumption of a Fermi acceleration of protons.
We show that Fermi acceleration of electrons is enough to account for the
gamma spectra, their shape, cut-off and their variability. Moreover the
spectral break is nicely explained by invoking an gamma-ray 
photosphere. Nevertheless we give estimates of the high energy cosmic ray
generation in AGNs and of the resultant neutrino flux, that turns out to
be very sensitive to the spectral index of the proton distribution.
\end{abstract}
\end{frontmatter}

\section{Introduction}

Active Galactic Nuclei are particular among the high energy engines for they
are both compact objects in a precise physical sense and the central power
engine of the most extended non thermal sources of the Universe. As such
they are sites of high energy electrodynamics with gamma ray and pair
production, and particle accelerators that could generate high energy
cosmic rays, that undergo hadronic processes and could emit high energy
neutrinos as a signature of this generation.

They are compact objects because they display an intense X-radiation
field that makes them optically thick to gamma-rays, at least within some
radius of order $100r_G$ ($r_G = 2GM_*/c^2$ is the gravitational radius
of the presumed black hole). This is measured by a number called
"compactness" and defined for a spherical region of size $R$ by:
\begin{equation}
        l \equiv \frac{\sigma_T L_X}{4\pi Rm_ec^3} \ .
        \label{CPS}
\end{equation}
where $L_X$ is the X-ray luminosity around 500 keV.\\
So when $l \gg 1$, every gamma-photon interacts with an X-photon to give
a pair of electron-positron. There are several possible processes that
produce gamma-photons and thus pairs in AGNs. They can be consequence of
the Penrose mechanism, the effect of the gap electric field in the
vicinity of the rotating black hole, of the Fermi processes that maintain
hadronic collisions (proton-proton collisions and proton-soft photon
collisions) and the Inverse Compton process.

The presumed central black holes are likely surrounded by an accretion
disk that have been devised to explain the intense black body emission
\cite{sha73}. Later the standard model has been modified to explain
the hard X emission (the so-called Comptonized disk model) and also to
account for jet production \cite{bla76,bla82} by assuming the
existence of opened magnetic field lines threading the disk. In fact,
no high energy physics can be figured out without a magnetic field
having a pressure at least in rough equipartition with the particle
pressure. This is another important aspect of the black hole accretion
disk to concentrate a magnetic field in large volumes especially when
the central mass is of order $10^8$ solar masses.  Jets, and
especially FR2 ones with their hot spots and extended lobes, are large
magnetic structures revealed by their powerful synchrotron radiation.
Fermi processes of first and second order need those large magnetized
regions to accelerate particles to very high energy. Indeed the size
of the accelerator determines the maximum energy through the maximum
gyro-radius it can contain and the transit time of the particle flow
in the acceleration region limits the acceleration period.

The purpose of the paper is to gathered the results we have obtained
recently on the high energy emission of AGNs and on particle acceleration
in order to explain why we prefer the electrodynamic explanation of the
gamma-ray emission of blazars and BL-Lac and to compare our pair model with
other electrodynamic models. Nevertheless we want to emphasize also the
interesting hadronic physics of AGNs (as proposed by \cite{pro,man})
and to discuss the possibility for AGNs to produce high energy cosmic rays
(between $10^6$ to $10^{11}$ GeV) and the consequent neutrino emission.

The paper is organized as follows. In section 2 we discuss the efficiency
of the Fermi processes. Section 3 is devoted to the discussion of the
nature of the physics that underlines the gamma emission of blazars
(hadronic or electrodynamic?) and the discussion of the various
electrodynamical models. The generation of high energy protons and
neutrinos in AGNs is examined in section 4.

\section{The efficiency of the Fermi processes}

It is often said that the first order Fermi process at shock is more efficient
than the second one, and that the Fermi processes accelerate protons more
efficiently. We will examine these issues. Then we will estimate the
highest energy achieved by electrons and protons in Active Galactic
Nuclei and their associated jets and extended regions.

\subsection{1st and 2nd order Fermi acceleration}

{\em Fermi acceleration at shocks}

A suprathermal particle that crosses a nonrelativistic shock front, comes
back up stream through pitch angle scattering and then crosses again the
shock front has gained an energy such that
\begin{equation}
        \frac{\delta p}{p} = \frac{4}{3}\frac{u_1-u_2}{v cos\theta_1} \ ,
        \label{DEP}
\end{equation}
where $\theta_1$ is the angle of the magnetic field line with respect
to the shock normal, and $u_1$ and $u_2$ are respectively the upstream
and downstream velocities of the flow. $v$ denotes the velocity of the
particle ($v \simeq c$ for a relativistic particle).\\
Its residence time in the vicinity of the shock is determined by the
downstream spatial diffusion coefficient $D$ through : $t_r =
2D/u_2^2$.  It thus depends on the main microscopic ingredient of the
theory namely the pitch angle frequency $\nu_s \equiv <\Delta
\alpha^2>/\Delta t$, where $<\Delta \alpha^2>$ is the mean quadratic
variation of the pitch angle during an interval $\Delta t$, larger
than the correlation time. In fact the diffusion coefficient depends
on the shock obliquity and one defines an effective diffusion
coefficient that combines the parallel and perpendicular coefficients,
$D_{\parallel} = \frac{1}{3}\frac{v^2}{\nu_s}$ and $D_{\perp}
=\frac{1}{4}r_L^2 \nu_s$, where $r_L=p/eB$ is the Larmor radius of the
particle. However the perpendicular diffusion coefficient likely
reaches the Bohm's value $D_B = \eta_0 r_L v$ with $\eta_0 \simeq 5
\times 10^{-2}$; its derivation for relativistic particles can be
found in \cite{ros}; so $D_{eff} = D_{\parallel} cos^2 \theta_2 +
D_{\perp} sin^2\theta_2$.  Except for almost perpendicular shocks
(which would be the most efficient for particle acceleration if Bohm's
diffusion is at work \cite{pelrol,jok}, the parallel diffusion
prevails.

At each crossing, the particle has a probability $\eta = 4u_2/v$ to escape.
For relativistic particles $v\simeq c$ and the crossing frequency is
almost $\nu_c = 1/(\eta t_r)$. The acceleration rate is therefore such that
\begin{equation}
        \frac{<\Delta p>}{\Delta t}=\frac{4}{3}\frac{u_1-u_2}{v cos\theta_1}
        \nu_c p = \frac{r-1}{3t_r}p \ .
        \label{ACT}
\end{equation}
For non-perpendicular shocks, the acceleration time scale $t_1$
of this first order Fermi process is thus
$t_1=t_r \sim (c^2/u_2^2) \nu_s^{-1}$. If the suprathermal particles would
couple with the thermal medium through ordinary collisions, the energy
exchange would lead to a thermalization of the fast particles. The
coupling with the thermal medium is not collisional, it is through the
magnetic disturbances (likely Alfven waves) of the thermal medium. Often
the Alfven velocity is smaller than the velocity of light and thus the
magnetic component of the Lorentz force is larger than the electric one
by a factor $v/V_A$. Therefore in first approximation, the interaction
with quasi static magnetic disturbances produces pitch angle scattering, which of course
does not lead to thermalization.
\par

{\em Stochastic acceleration}

Alfven waves in a turbulent plasma produce not only a pitch angle
scattering but also an energy diffusion which is of second order in
$V_A/c$ and is also a second order process in term of Fokker-Planck
description. One has
\begin{equation}
        \frac{<\Delta p^2>}{2\Delta t} \equiv \frac{p^2}{2t_2} \sim
        \frac{V_A^2}{c^2} \nu_s p^2 \ .
        \label{ACT2}
\end{equation}
This defines the characteristic time $t_2$ of a second order Fermi
type acceleration process, which does not tend to thermalize either.
Since the downstream flow after a shock has a subsonic velocity $u_2$,
for a magnetic field at rough equipartition, $u_2 \sim V_A$. It means
clearly, as recognized by Jones \cite{jon} and somehow by Campenau and
Schlickheiser \cite{cam}, that the second order process is not less
efficient than the first order Fermi process at shock, since both
times are of order $(c^2/V_A^2) \nu_s^{-1}$ and are controlled by the
same pitch angle scattering frequency.
\subsection{Acceleration time scale and maximum energy}

The pitch angle scattering frequency depends only on the momentum
of the particle and the detailed dependence is determined by the Alfven
wave spectrum. Only particles having a Larmor radius comparable to a
wavelength of the spectrum undergo scattering. In particular, the minimum
momentum ($p_0=  m_pV_A$) for interaction is determined by the smallest
wavelength $\lambda_0 = 2\pi V_A/\omega_{cp}$ 
($\omega_{cp}$ is the cyclotron pulsation of the non relativistic
protons) and the
maximum momentum achievable by Fermi acceleration is given by the maximum
wavelength ($p_m = p_0\lambda_m/\lambda_0$). The momentum dependence of
the pitch angle scattering frequency is sketched in fig.1, for an Alfven
spectrum in $\omega^{-\beta}$, $\nu_s \propto p^{\beta-2}$ between $p_0$
and $p_m$.

\begin{figure}[h]
\begin{center}
\rotatebox{0}{
\resizebox{120mm}{100mm}{
{\includegraphics{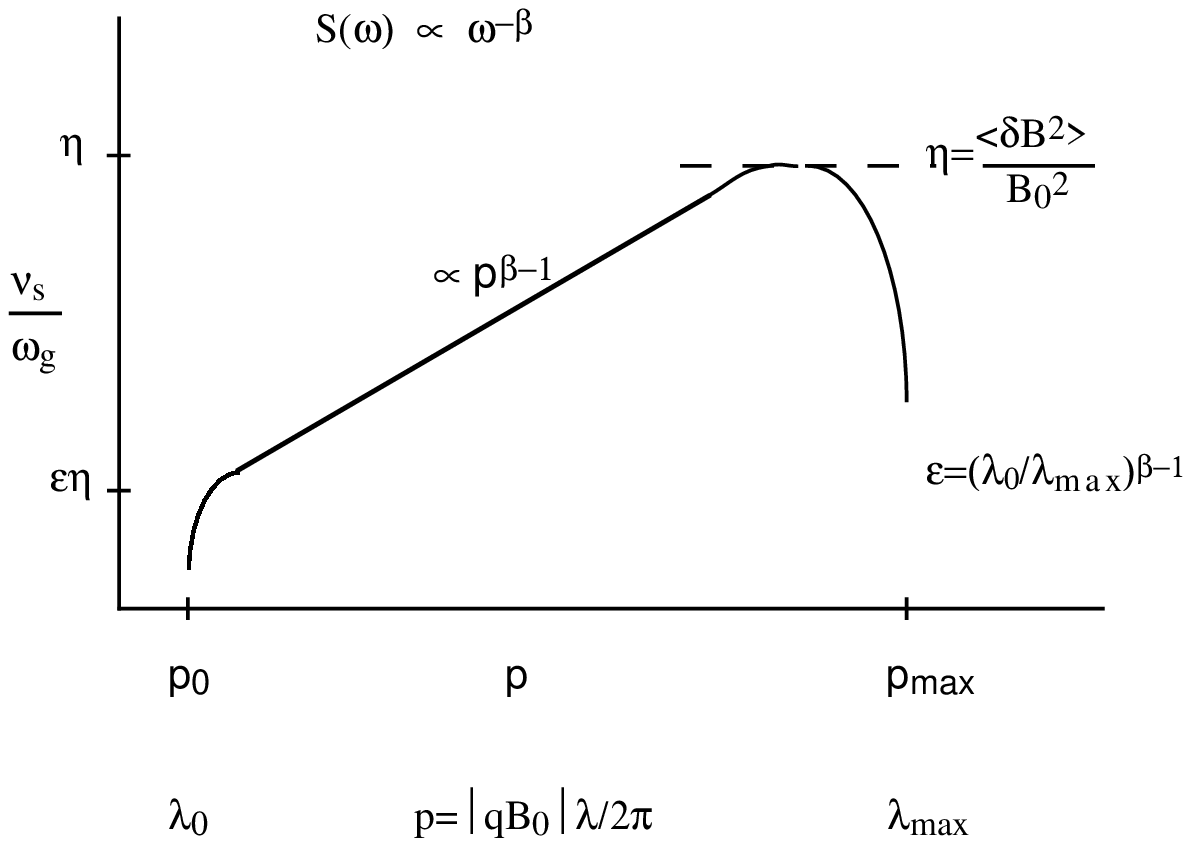}}}}
\caption{The momentum dependence of
the pitch angle scattering frequency. 
Magnetic perturbations scatter preferencially particles
having a Larmor radius close to the wavelength of Fourier mode. The
minimum wavelength determines the threshold for scattering and acceleration,
and the maximum wavelength determines the maximum energy for the
particles to be scattered and accelerated. The efficiency of the
scattering and of the acceleration at a given energy depends on the amplitude
of the (nearly) resonant mode, the Alfven spectrum being in
$\omega^{-\beta}$. Precise coefficients can be calculated in quasilinear 
theory valid for a turbulence level $\eta \ll 1$. However the estimate is still roughly correct for $\eta \sim 1$.}
\end{center}
\protect\label{fig1}
\end{figure}

What are the best conditions to get a fast acceleration? Clearly the
fastest acceleration process takes place in a relativistic plasma that
has an Alfven velocity close to the velocity of light. The modified
Alfven velocity in a relativistic plasma is given by:
\begin{equation}
        V_{A,rel} = \frac{c}{\sqrt{1+2\frac{P}{P_m}}} \ ,
        \label{VM}
\end{equation}
where $P$ is the relativistic pressure and $P_m$ is the magnetic pressure.
At equipartition the modified Alfven velocity equals the relativistic
sound velocity $c/\sqrt{3}$. Since these plasmas are supposed to be
magnetically confined the propagation velocity of the electromagnetic waves
is close to the velocity of light. Under those conditions, pitch angle
scattering and acceleration works with the same time scale and the usual
expansion in power of $V_A/v$ cannot be done. Moreover
the second order Fermi process is efficient and one does not
know whether the first order Fermi process works at relativistic shocks.

{\em Electron relativistic plasmas}

As we just saw, for high energy particles, the only
efficient scattering process comes from the resonant interaction of these
particles with Alfven waves, which occurs with the waves having a
wavelength almost equal to the Larmor radius of the scattered particle.
In ordinary plasmas, the most massive component is due to non relativistic
protons and the
Alfven waves develop at wavelengthes larger than $V_A/\omega_{cp}$.
This puts a severe threshold for resonant interaction, especially for the
electrons that must be very energetic already: $p_0 > m_p V_A$; the
corresponding threshold energy $\epsilon_0 \simeq p_0c$ is of order one
MeV in supernovae remnants and often of order 100 MeV in
extragalactic jets. However all
the single charged relativistic particles having the same momentum are
accelerated in the same way by the Fermi processes. Saying that protons
are accelerated more efficiently than electrons is not true. The only
trouble in ordinary plasmas is that protons are more numerous above the
resonance threshold and the electrons must be efficiently injected
above the threshold to participate to the Fermi processes. Several
processes are known to inject electrons in the cosmic ray population,
such as the development of a parallel electric field component in
magnetic reconnections, or short waves (magnetosonic or whistler)
\cite{rag}.

In compact objects, "exotic" plasmas can be created with a copious
relativistic electron (positron) component. The cauldron of the black
hole environment \cite{hen1,mar2} could likely be dominated by the
pair plasma. When the most massive component is due to relativistic
electrons (positrons), they are more numerous above the resonant
threshold. Under these interesting conditions, the power of the
acceleration process goes almost entirely in the radiative particles,
which is the best regime to have the most efficient conversion of
energy into radiation.

These exotic plasmas (either simply relativistic electrons dominated
with non relativistic protons or pair dominated) have interesting
dynamics.  First, they can be propelled at relativistic velocities by
the Compton rocket effect provided that they are maintained hot in the
cauldron \cite{hen1,ren98}.  Second, the investigation of the
nonlinear regime of Alfven disturbances \cite{pel} shows that
acceleration works efficiently only when the magnetic pressure is
larger than the plasma pressure. Overpressured plasmas (not confined)
suffer radiative cooling and thus come back to rough equipartition.
The nonlinear development of such waves gives rise to intense
relativistic fronts that accelerate particle more efficiently than
through familiar Fermi processes.  The average acceleration time scale
can easily be as short as hundred gyro-periods.

For a first investigation, we will express the efficiency of the
acceleration process by a factor $A$ that measures the acceleration time
in term of the gyro-period:
\begin{equation}
        t_a = A T_g(\gamma)
        \label{TOA}
\end{equation}
(some authors take $A =10$, or even one! this is not reasonable and
unconsistent with the observed synchrotron emission). The acceleration
factor $A$ depends on the shock obliquity (if any), of the turbulence spectrum,
of the energy of the particle. In the case of shock acceleration, it
reaches a minimum value for nearly perpendicular but still subluminal
shocks, if the Bohm's diffusion prevails downstream. Then $A \simeq 0.5
c^2/u_2^2$, but the crossing number ($c/4u_2$) must remain large for the
theory to apply; typically $u_2$ must remain smaller than $0.1c$, which
puts $A$ significantly larger than $10$ in the best conditions.

At a given momentum, electrons, positrons and protons are accelerated at the
same rate, only loss processes make difference in the shape of their
spectra by introducing a cut-off. Their number densities differ because of the
threshold defined by the smallest momentum $p_0$. In ordinary plasma, like in
the interstellar medium, protons can reach more easily the threshold momentum
than electrons, that is probably why they are more numerous in the galaxy
cosmic ray population. The synchrotron loss time for electrons and positrons is
\begin{equation}
        t_{rad}(\gamma) \simeq 0.75\frac{10^9}{\gamma}
\left(\frac{1{\rm G}}{B}\right)^2 {\rm sec.}
        \label{TSY}
\end{equation}
Electrons suffer Inverse Compton losses in the nuclei also; 
however assuming that
the soft photon energy density is not much larger than the magnetic
energy density, the synchrotron time given by eq.(\ref{TSY}) is a correct
estimate of the radiation time scale. The maximum energy achievable by the
acceleration process is obtained by setting
$t_a(\gamma_{me}) = t_{rad}(\gamma_m)$; thus
\begin{equation}
        \gamma_m \simeq 4.6 \times \frac{10^7}{\sqrt{A}} 
\left(\frac{1{\rm G}}{B}\right)^{1/2} \ .
        \label{GAM}
\end{equation}

The energy of protons is limited by two conditions: the size of the
accelerator and the residence time of the proton in the accelerator.
These two conditions are roughly numerically equivalent to the more accurate
conditions: the larger size of the magnetic perturbations and the
characteristic time of variability of the source.
Independently of the efficiency of the acceleration process, the maximum
energy achievable by protons in an accelerator of size $R$ is:
\begin{equation}
        \gamma_{mp} \simeq 10^{12} \frac{R}{1{\rm pc}}\frac{B}{1{\rm G}} \ .
        \label{GAMS}
\end{equation}
However the most severe constraint generally comes from the residence time 
$t_{res}$:
\begin{equation}
        \gamma_{mp} \simeq 1.5 \times \frac{10^3}{A} \frac{t_{res}}{1 {\rm sec}}
        \frac{B}{1{\rm G}} \ .
        \label{GAMT}
\end{equation}
We remark that the maximum energy depends on the product of the magnetic
field intensity by a size (the residence time is proportional to the size
of the accelerator). This tends to favor the vicinity of the black hole
to achieve the highest energy, because this product is likely decreasing
with the distance to the black hole.

\section{The gamma-ray emission of Blazars}

\subsection{Observational constraints}

Non-thermal electromagnetic emission, i.e. synchrotron and high energy
radiation, is so far the only direct evidence for the existence of
highly relativistic particles in AGN. The high energy cut-off and
variability timescale of its spectrum provide thus important
constraints on the size and the physical characteristics of the
accelerating region.  All radio-loud objects are characterized by an
intense non-thermal synchrotron emission ranging from radio to
optical, and sometimes hard X-ray range. The corresponding high energy
component extends from soft X-ray to hard gamma-ray range. In leptonic
models, it is interpreted as the Inverse Compton scattering of soft
photons, either the synchrotron or the thermal emission from an
external accretion disk. In hadronic models, it is the result of pair
cascades resulting from initial photopion processes. In any case,
observation of photons with an energy $\gamma m_{\rm e}c^2$ requires
particles with a Lorentz factor at least $\gamma$.\\
The high energy cut-off is not well measured by the current
observations.  Most gamma-ray emitting AGNs seen by EGRET up to 30 GeV
have not been detected by higher energy ground-based Cerenkov
telescopes, while the extrapolation of the EGRET spectra should have
often fallen above their sensitivity limit. The lack of available
instruments in the 30 GeV-300 GeV range has up to now prevented us to
better determine the position of the high energy cut-off. The actual
starting of improved sensitivity Cerenkov telescopes like the CELESTE
project should help to fix this issue. However, interpretation of the
cut-off can be complicated by the extragalactic absorption on the
IR-optical-UV background \cite{sal}.  Assuming a cut-off around 50
GeV, a conservative estimate of the maximal Lorentz factor is thus
about $10^{5}$. Flares on timescales of some days have often been
observed in radio-loud quasars such as 3C279 and 3C273, which gives a
size of $\sim 10^{16}$ cm for a static isotropic emission zone. With
the huge luminosity observed during the flares, this should produce
the complete absorption of gamma-rays by pair production. As we
discuss in the next sections, this paradox can only be solved by
invoking an emission zone with a relativistic bulk motion in a
direction close to the observer's
line of sight. \\
Three objects however have already been detected by Cerenkov
telescopes above 200 GeV: Mrk 421, Mrk 501 and 1ES 2344+51.4. They all
belong to the class of BL Lacs, and are relatively close objects with
an intrinsic relatively low luminosity. As a class, BL Lacs are
characterized by the weakness of the thermal component and of their
emission lines (if any). The synchrotron component power is often
comparable to the high energy one. There seems to be also a strong
correlation between the frequency of the maximum of the synchrotron
component and that of the high energy component. Indeed, all TeV
blazars are X-ray BL Lacs, for which the synchrotron component extends
also to the hard X-ray range. In the case of Mrk 501, a spectacular
synchrotron flare has been observed by Beppo SAX up to 100 keV, with a
corresponding simultaneous TeV flare \cite{Pia98}. In the leptonic
model, it is thus plausible that the entire spectrum is mainly
produced by SSC process, with a variable maximum particle Lorentz
factor and/or magnetic field. Observation of photons up to 10 TeV
imply the existence of particles with $\gamma_{max} \ge 2\ 10^{7}$.
These particles can produce synchrotron radiation up to 100 keV provided that
magnetic field $B \simeq 0.02 {\rm G}$. Moreover, flares are observed
with timescales less than one hour, which implies a accelerating
region as small as $10^{14}$ cm for a static source.  However, as we
discuss below, all these estimates must again take into account the
bulk motion of the relativistic jet which emits this high energy
radiation.

\subsection{Hadronic versus electrodynamic processes}

It turns out that fast variability is more likely related to the
variation in the acceleration process rather than in photons burst
\cite{hen3}. So to know whether the underlying physics of
gamma-emission is of hadronic or electrodynamic nature, we look at the
characteristic time for particle acceleration in the region where the
gamma source is located. This source is likely located around $100r_G$
at the beginning of the jet \cite{hen2,mar1}.

For electrons at the gamma-ray photosphere, located around $100 r_G$
(a day-light, more or less), where presumably most of the
gamma-emission takes place,
a magnetic field of 10 G leads to $\gamma_{me} \sim 10^6$ within 10 sec.
with $A = 10^3$.

For protons at the gamma-ray photosphere, they could reach $10^{10}$ GeV, but
in 10 years with $A = 10^3$, whereas the variability time scale and the
transit time for a jet portion of $100r_G$ is not more than a day. This
reduces the energy to $10^6$ GeV.

\subsection{The blazar spectra and variability}

If emitted by a spherical static source, the gamma radiation observed 
in blazars could not escape from the central
region because of its compactness. However the gamma emission is
observed only in some blazars \cite{mon} that have jets with
superluminal motions and is understood if one takes into account the
Doppler beaming. Indeed the relativistic motion (of bulk velocity
$\beta_b c$ of corresponding bulk Lorentz factor $\gamma_b\equiv
1/\sqrt{1-\beta_b^2}$) of the emitting cloud along a direction having
a rather small angle $\theta$ with respect to the line of sight
produces relativistic aberrations that depends on the Doppler factor
$\delta$:
\begin{equation}
        \delta \equiv \frac{1}{\gamma_b(1-\beta_b cos \theta)}
        \label{DF}
\end{equation}
Superluminal motions indicate that the Doppler factor can be 
as large as 10-20 and thus
the radiation is beamed towards the observer according to:
\begin{equation}
        I(\omega, \theta) = \delta^3 I_0(\frac{\omega}{\delta}) \ ,
        \label{AL}
\end{equation}
thus the luminosity appears much larger than its intrinsic value which, in
fact, remains smaller than the UV bump. Moreover the variability time
scale appears shorter by a Doppler factor:
\begin{equation}
        \tau_{obs} = \frac{\tau}{\delta} \ .
        \label{TOBS}
\end{equation}
Since the variability time
gives the maximum size of the source, the gamma emission region is not
larger than $100 r_G$.

Thus the natural understanding of the high energy emission of blazars
is to consider that it comes from the region where relativistic clouds
become optically thin to gamma rays at about $100 r_G$ \cite{hen2}.  A
detailed calculation of the radiative transfer of the high energy
photons (X and gamma) emitted by the electrons of a relativistic jet
in the anisotropic radiation field of an accretion disk has been
performed \cite{mar1}. It takes into account the Inverse Compton
process on the UV bump and the pair creation process by two gamma
photons, the X- and $\gamma$-rays being generated by the Inverse
Compton process and there is also a small contribution of the
annihilation. Because of the stratification and the growth of the pair
density up to a value that makes the source optically thick to Thomson
scattering of the soft photons, the spectrum breaks around few MeV,
the higher $\gamma$-rays being still in the optically thick regime,
whereas the X-rays are optically thin to $\gamma \gamma$-pair
production. This model is the only one that accounts for the observed
spectrum break, since it predicts that the gamma index is twice the X
index, whereas the incomplete Compton cooling model \cite{der}
predicts the canonical $1/2$ inflection.

\subsection{Comparison of the electrodynamics models}
\par

In these models, the high energy photons come from
the comptonisation of various sources
of soft photons by relativistic electrons or electron-positron pairs.
 The leptonic plasma is assumed to be
isotropic in a blob moving relativistically away from the central source.
\par
{\em Models with no pair creation}
\par
In the model of Dermer \& Schlickeiser \cite{der}, the soft photons
source is the accretion disk radiation. The high energy emission takes
place in a small angle $\sim 1/\gamma_b$ from the direction of the jet
axis with a bulk Lorentz factor set to a constant (typically $\gamma_b
\simeq 10$).  The spectral break (0.5) is due to incomplete cooling of
the electrons distribution.  Electrons are just injected, and there is
no pair creation. All the energy are emitted in the same region which
implies that simultaneous variation should be observed in all
wavelength.
It was argued \cite{sik94} that this model could not avoid pair
creation unless the source of gamma-ray photons is located at
distances $\geq 10^{17}$ cm from the central black hole. At this
distance the direct disk radiation is strongly redshifted, and Sikora
et {\it al.}  proposed that the dominant contribution to the soft
photons density energy is the rescattered radiation from BLR clouds.
In their model, they took the same assumptions as Dermer \&
Schlickeiser for the description of the electronic plasma and they
also considered that all energy photons are emitted in the same
region.

Inhomogeneous Synchrotron Self-Compton emission were proposed
\cite{ghi} to explain BL-Lac spectrum where there is no evidence of
accretion disk nor lines in the observations. The soft photons are
synchrotron photons emitted in the jet and the resultant spectrum is a
sum of local synchrotron and comptonization emission in the
inhomogeneous jet.  This model predicts that flux variation is greater
at high energy than at low ones. A spectral break can be explain by
the inhomogeneous geometry of the jet, and can be different from 0.5.
Nevertheless the model has the so-called "Inverse Compton Catastrophe"
problem close to the nucleus \cite{der2}. However it seems to work
nicely for BL-Lacs.
\par

{\em Models with pair creation}
\par


Blandford and Levinson \cite{bla} combined Inverse Compton on
UV-photons from the disk, rescattering radiation or jet itself.
They took into account opacity to
 pair creation on the soft photons and solved numerically the kinetic
equations for the different
 populations in the comoving frame with constant bulk Lorentz factor
($\gamma_b=10$). The electron-positron pairs are injected to some threshold
corresponding to acceleration efficiency. The emergent X-ray to gamma-ray
spectrum exhibits a spectral break due to pair opacity effect. The breaking
energy and the index variation depend on the spectrum and radial variation
of the soft radiation. In this model X-ray and submm to optical emission
 originates from region 100 times closer to the central engine than gamma-ray
 emission and $10^4$ times closer than radio emission. One can then predicts
 the evolution time of an observed flare.
 
 In the Henri et al. approach \cite{hen1,hen2,mar1}, the
 reacceleration process in an important aspect of the model. It
 maintains the source in a regime close to the pair-creation
 catastrophe, which is interesting to explain the fast variability.
 Moreover for a reheated pair plasma, the radiation pressure from the
 accretion disk due to Compton interaction (`Compton rocket effect')
 can gradually accelerate the plasma along the jet to high enough bulk
 Lorentz factor, which is not a free parameter in this model. They
 considered Compton interactions of a relativistic pair plasma on soft
 photons from an accretion disk.  The pair plasma is created in the
 vicinity of the central black hole where the opacity to pair creation
 is greater than one.  When the soft photon population is depleted by
 Compton interactions, the jet becomes optically thin to gamma photons
 which can escape.  The dependence on energy of the photosphere
 explains the spectral break of the gamma spectrum around a few MeV.
 Indeed one has $\alpha_{\gamma} = 2\alpha_X$. The model predicts
 spectrum in good agreement with observation for 3C273, 3C279, and
 CenA. As in Blandford \& Levinson model, due to pair opacity effects,
 X rays should arise before gamma-rays during a flare sequence, but
 with a much shorter delay. Nevertheless for a complete
 electrodynamic model, one needs to take into account Compton
 interactions on both UV-photons and synchrotron photons together with
 pair creation.
 
 A very recent publication \cite{ward98} gives a strong argument in
 favour of a pair model; this is based on the polarization measurement
 in the jet of 3C279.

\section{Protons and hadronic processes}

The extragalactic origin of high energy cosmic rays, beyond $10^{14}eV$,
is very likely. Active Galactic Nuclei and their extended regions are
the most obvious candidates as being the acceleration sites. The other
candidates are the Gamma Ray Bursts and the Topological Defects. The
present knowledge of AGNs allows to give better estimates of the
parameters to test this idea. High energy protons, especially beyond the
threshold for the so-called GZK effect, produce gamma rays and high
energy neutrinos through collisions with soft photons. The idea that the
ultimate energy cosmic rays are produced in AGNs with a signature with
the gamma rays and the neutrino emission is really appealing.
However we show in the previous section that the gamma-ray emission of
blazars is more likely explained by purely electrodynamical processes
with Fermi acceleration of electrons. The nucleus that could be a source
of gamma-rays is optically thick to them because of the pair production
process. So we think that the gamma spectrum and the possible neutrino
spectrum are not correlated. Although the gamma emission is more likely
explained by electrodynamics, the protons acceleration and neutrino
emission are likely to occurs in AGNs as well. We will present now our
estimates of protons energy and neutrino flux.

\subsection{The cosmic ray production in the AGN components}

The two main regions of particle acceleration are the environment of the
black hole and the hot spots (in case of FR2 jets).

Let us consider first
the electrons.

i) Within $10 r_G$ in the vicinity of the AGN black hole, a magnetic field
of order 1 kG (equipartition) can be concentrated. The electrons can
therefore reach $\gamma_{me} \sim 10^5$ in $10^{-2}$ sec. with $A = 10^3$.

ii) In jet hot spots like those of Cygnus A, a typical value of the magnetic
field is $B= 10^{-4}$ G; which gives $\gamma_{me} \simeq 4 \times 10^8$
within a time of $10^8$ sec. This value is a little too high, as
compared to the synchrotron data; which means that the efficiency of the
acceleration process, implied by chosing $A= 10^3$, is a little overestimated.

Let us consider the maximum energy that the protons can reach with the
same assumption expressed by eq.(\ref{TOA}) with $A =10^3$ for the acceleration
process.
There are two limits: one is implied by the maximum MHD scale, the other by
residence time of the proton in the accelerator.

i) In the vicinity of the black hole, within $10 r_G$, the protons could
reach $10^{10}$ GeV within a year if they would be confined. However for
a flow of velocity $c/10$, their residence time would fall to $10^5$ sec.,
which would reduce their maximum energy to $10^8$ GeV.

ii) In jet hot spots of size of order 1 kpc, the protons could reach
$10^{11}$ GeV within $10^7$ years, which is comparable with the age of the
source. But they could spend much less time in the hot spots; if the
downstream flow has a velocity of $c/10$, they stay only $10^4$ years,
which lower the maximum energy to $10^8$ GeV.

Only a significant increase of the efficiency of the acceleration
process in favour of the protons, since this is not needed for the
electrons, could compensate the shortness of the residence time in the
nucleus and in the hot spot. But it is not reasonable to expect much
better than a factor $10$, since even for big magnetic disturbances,
the time for pitch angle scattering is always longer than the
gyroperiod and the Fermi acceleration time is always longer than the
scattering time. Alternately, leaving the nucleus with $10^8$ GeV they
could be reaccelerated all along the jet and reach $10^{11}$ GeV in
the hot spot and then escape in the extended lobes. If so, the
extragalactic jets could be the accelerators of the high energy cosmic
rays \cite{pro,man}.

The maximum energy achievable by the protons in an AGN is therefore:
\begin{equation}
        \epsilon_{max} \simeq \frac{M_*}{10^8 M_o}\frac{B}{{\rm 1kG}}\frac{10^{20}}{A}
        \ {\rm eV} \ .
        \label{EMAX}
\end{equation}
Nevertheless these estimates do not convince that AGNs definitely are the
sources of the highest energy cosmic rays, since $A > 10$ ({\it cf} comments after equation (6)). Should we consider large
structures like collisions of galaxies with shock fronts of Mpc size,
but with magnetic fields as weak as $\mu$ Gauss?

\subsection{gamma ray and neutrino emission}

{\em Neutrino emission from the nucleus}
\par

High energy neutrinos (energy larger than $100$ MeV) can be emitted by
pp-collisions, with a cross section $\sigma \simeq 2.7 \times
10^{-26}{\rm cm}^2$ for proton energy larger than 2 GeV. 
The neutrino luminosity depends on the density of protons of momentum
larger than $p$:
\begin{equation}
        n(>p) = \chi n_*\left(\frac{p}{p_0}\right)^{1-\eta} \ ,
\end{equation}
where $n_*$ is the particle density in the disk and $\chi$ is the 
fraction of proton number above the threshold $p_0$.
Assuming $n_* = 10^{14} {\rm cm}^{-3}$, it turns out that the direct pp-process is 
always very tiny.

The photo-production of pions is expected to be more efficient to produce
neutrinos through the $\Delta-$resonance (the so-called GZK effect)\cite{pro}:
\begin{eqnarray}
         &  & \stackrel{(2/3)}{\rightarrow}  p + \pi^0 \rightarrow p +
         \gamma + \gamma   \nonumber \\
        p + \gamma & \rightarrow  \Delta^+   &  \label{GZK} \\
         &   & \stackrel{(1/3)}{\rightarrow}  n + \pi^+ \rightarrow ...
        \rightarrow p + e^+ + e^- + \nu_e + \bar \nu_e + \nu_{\mu} +
        \bar \nu_{\mu}  \nonumber
\end{eqnarray}

For a head-on collision, the threshold energy of the proton is
\begin{equation}
        \epsilon_{th} = \frac{m_{\Delta}^2 - m_p^2}{4 \epsilon_{\gamma}} \ ;
        \label{SGZK}
\end{equation}
and the cross section
$\sigma_{p\gamma} \simeq 5.4 \times 10^{-28} {\rm cm}^2$.
The process can occur in the AGNs with the UV-photons where the threshold
energy is of order $10^{16}$ eV, since we saw in the
previous subsection that the protons can reach much higher energies.

A neutrino emission is thus possible and
\begin{equation}
        L_{\nu} = \int \int \epsilon_{\nu}n_{ph} \sigma_{p\gamma} c f_p d^3pd^3V
        \simeq \frac{\bar \epsilon_{\nu}}{\bar \epsilon_{ph}}L_{UV}
        \sigma_{p\gamma}\frac{R}{3} n_p(>\gamma_{th}) \ .
        \label{LEX}
\end{equation}
We have got the following estimate:
\begin{equation}
        \frac{L_{\nu}}{L_{UV}} \sim 10^{14-8\eta} \ .
        \label{LNUG}
\end{equation}
For $\eta = 2$ there are enough protons above the high threshold
($\gamma_{th} \sim 10^7$) to have $L_{\nu} \sim 10^{-2}L_{UV}$, whereas for
$\eta = 3$ the ratio is only $10^{-10}$ ! 
$\eta = 2$ is expected in the vicinity of shocks. However after having traveled over a distance that makes them sensitive to diffusion or losses the integrated cosmic rays distribution decays from
$\eta = 2$ to $\eta = 2.7-3.1$.

The GZK-effect occurs also during the propagation of the cosmic rays
in the intergalactic medium where they collide with the cosmological
black body. The threshold is of order $10^{20}$ eV and cosmic rays of
larger energy cannot come from sources beyond $100$ Mpc \cite{aha}.

\section{Discussion}

Does the gamma-emission of blazars allow to discriminate whether
the underlying physics is of electrodynamic or hadronic nature?
This discussion focuses on the two issues of acceleration and variability.
Of course the main argument in favor of the electrodynamic model is that
the electrons allow a much faster variability than protons. The small
size of the high energy sources revealed by their variability would imply
a strong magnetic field to have proton Larmor radii smaller than the size.
It is often unduly said that the electrons are not efficiently
accelerated by Fermi processes, that shocks accelerate more
efficiently than the second order Fermi process, and also that they accelerate
mostly protons. It has been shown that these prejudices are not plainly
true.

The analysis of the second order Fermi acceleration of the relativistic
electrons does not reveal any serious difficulty to explain the gamma
emission of blazars, even to explain the few TeV radiation of the BL-Lacs
(Mrk 421, Mkr 501 and maybe 1ES 2344+514). The Klein-Nishina limit seems to be the major
limitation of the inverse Compton emission on accretion disk photons,
the cut-off should be at higher energy for the Synchrotron Self-Compton
emission. Thus the emission of BL-Lacs beyond TeV energy is likely
the SSC-radiation.

In the case of quasars, the inverse Compton process can also be
accompanied by the pair creation process. This seems in fact
unavoidable within $100 r_G$, and it could explain nicely the spectrum
break around few MeV. Indeed only the pair model \cite{mar1} was able
to explain so far a gamma-spectrum index which is twice the X-spectrum
index as observed.

-Are AGNs the sources of the high energy cosmic rays? They are
certainly sources of high energy cosmic rays but the possibility to
reach $10^{20}eV$ is difficult with the usual Fermi processes.
Acceleration in relativistic plasmas seems promising \cite{pel}.

-Are they localized sources of neutrinos? Yes certainly, but the flux
would be measurable only if the cosmic ray index would be close to $2$ in the
source.

-Are the neutrino emissions correlated with the gamma-ray emission? The
GZK-effect suggests that as many neutrinos as gamma photons are emitted;
however we saw that gamma-photons cannot escape from the black hole
environment; they come from the gamma-ray photosphere. Moreover we argued
that the gamma spectrum is very likely explained by purely
electrodynamical processes maintained by Fermi acceleration of electrons.
Thus we think that the neutrino emission is not correlated with the gamma
emission.


\begin{thebibliography}{99}

\bibitem{aha}
Aharonian, F. A., \& Cronin, J. W., 1994, Phys. Rev. D, 50, 1892.

\bibitem{bla76} 
Blandford, R. D. \& Znajek, R. L., 1977, MNRAS, 179, 433.
    
\bibitem{bla82} 
Blandford, R. D. \& Payne, D. G., 1982, MNRAS, 199, 883. 

\bibitem{bla}
Blanford, R. D., \& Levinson, D., 1995, ApJ, 441, 79.
  
\bibitem{cam}
  Schlickeiser, R., Campeanu, A., \& Lerche, L., 1993, A\&A, 276, 614.
  
\bibitem{der}
  Dermer, C. D., \& Schlickeiser, R., 1993, ApJ, 416, 458.
  
\bibitem{der2}
  Dermer, C.D., Sturmer, S.J., \& Schlickeiser, R., 1997, ApJS, 109, 103.
  
\bibitem{ghi}
  Ghisellini, G., George, I.M., Fabian, A.C., \& Done, C., 1991, MNRAS, 248, 14.
  
\bibitem{hen1}
  Henri, G., \& Pelletier, G., 1991, ApJ, 383, L7.
  
\bibitem{hen2}
  Henri, G., Pelletier, G., \& Roland, J., 1993, ApJ, 404, L41.

\bibitem{hen3}
  Henri, G., \& Renaud, N., 1998, in preparation.
  
\bibitem{jok}
  Jokipii, J.R., 1987, ApJ, 313, 842.
  
\bibitem{jon}
  Jones, F., 1994, ApJS, 90, 561.
  
\bibitem{lic}
  Lichti, G. G.,  et al., 1995, A\&A, 298, 711.
  
\bibitem{man}
  Mannheim, K., \& Biermann P., 1992, A\&A, 253, L21.
  
\bibitem{mar1}
  Marcowith, A., Henri, G., \& Pelletier, G., 1995, MNRAS, 277, 681.
  
\bibitem{mar2}
  Marcowith, A., Pelletier, G., \& Henri, G., 1997, A\&A, 323, 271.

\bibitem{mon}
  Von Montigny, C., et al., 1995, ApJ, 440, 525.
  
\bibitem{pel}
  Pelletier, G., \& Marcowith, A., 1998, ApJ, 502, 598.
  
\bibitem{pelrol}
  Pelletier, G., \& Roland, J., 1986, A\&A, 163, 9.

\bibitem{Pia98}
  Pian et al., 1998, ApJ, 492, L17. 

\bibitem{pro} Protheroe, R.J., \& Stanev, T., 1992, in High Energy
  Neutrino Astronomy, eds.  V.J. Stenger et al. (World Scientific,
  Singapore), p. 40.
  
\bibitem{rag}
  Ragot, B.,\& Schlickheiser, R., 1998, APh, 9, 79.
 
\bibitem{ren98}
  Renaud, N., \& Henri, G., 1998, MNRAS, 300, 1047.

\bibitem{ros}
  Rosso, F., \& Pelletier, G., 1993, A\&A, 270, 416.
  
\bibitem{sal}
  Salamon, M., \& Stecker, F. W., 1998, ApJ, 493, 547.

\bibitem{sha73}
  Shakura, N. I., \& Sunyaev, R. A., 1973, A\&A, 24, 337.
  
\bibitem{sik94}
  Sikora, M., Begelman, M., \& Rees, M., 1994, ApJ, 421, 153.
  
\bibitem{ward98}
 Wardle, J. F. C., Homan, D. C., Ojha, R., Roberts, D. H., 1998, Nature, 395, 457
  
\end{thebibliography}
\end{document}